\documentclass{article}

\usepackage{arxiv}

\usepackage[utf8]{inputenc} 
\usepackage[T1]{fontenc}    
\usepackage{hyperref}       
\usepackage{url}            
\usepackage{booktabs}       
\usepackage{amsfonts}       
\usepackage{nicefrac}       
\usepackage{microtype}      
\usepackage{lipsum}
\usepackage{graphicx}
\graphicspath{ {./images/} }

\title{Restoration of the JPEG Maximum Lossy Compressed Face Images with Hourglass Block based on Early Stopping Discriminator}

\author{
 Jongwook Si \\
  Dept. Computer AI Convergence Engineering\\
  Kumoh National Institute of Technology\\
  Gumi, KOREA 39177 \\
  \texttt{jwsi425@kumoh.ac.kr} \\
   \And
 Sungyoung Kim \\
  Dept. Computer Engineering\\
  Kumoh National Institute of Technology\\
  Gumi, KOREA 39177 \\
  \texttt{sykim@kumoh.ac.kr} \\
  \And
}

\begin{document}
\maketitle
\begin{abstract}
When a JPEG image is compressed using the loss compression method with a high compression rate, a blocking phenomenon can occur in the image, making it necessary to restore the image to its original quality. In particular, restoring compressed images that are unrecognizable presents an innovative challenge. Therefore, this paper aims to address the restoration of JPEG images that have suffered significant loss due to maximum compression using a GAN-based net-work method. The generator in this network is based on the U-Net architecture and features a newly presented hourglass structure that can preserve the characteristics of deep layers. Additionally, the network incorporates two loss functions, LF Loss and HF Loss, to generate natural and high-performance images. HF Loss uses a pretrained VGG-16 network and is configured using a specific layer that best represents features, which can enhance performance for the high-frequency region. LF Loss, on the other hand, is used to handle the low-frequency region. These two loss functions facilitate the generation of images by the generator that can deceive the discriminator while accurately generating both high and low-frequency regions. The results show that the blocking phenomenon in lost compressed images was removed, and recognizable identities were generated. This study represents a significant improvement over previous research in terms of image restoration performance.
\end{abstract}

\section{Introduction}
The image format can be used to store and represent image data, and it can be categorized into two main types: raster and vector methods. The raster method involves representing an image through pixels, including color information, while the vector method uses an equation to represent an image. The quality of the raster image increases with the number of pixels, but this also increases the file size. Enlarging or reducing the raster image results in a loss of image quality. In contrast, vector images remain clear even when enlarged, but it is challenging to express detailed information or natural changes due to the large number of calculations involved. JPEG, GIF, and BMP files use the raster method, while AI and SVG files use the vector method. Among these image formats, this study focuses on the JPEG image, which uses the raster method.\\
JPEG images utilize loss compression technology to intentionally remove certain image data and reduce file size. Only the parts that are not relevant to human vision are removed during this process. The compression process for JPEG images involves several steps. First, the color space is converted to eliminate irrelevant image content. RGB images are converted into YCbCr color space since humans are more sensitive to brightness components. Cb and Cr are compressed while maintaining Y. Next, 4:2:0 subsampling is performed to further reduce file size. After subsampling, Dis-crete Cosine Transform (DCT) is applied to each block, which divides the image into 8x8 blocks to express the high frequency component representing fine details. The coefficients are then quantized to remove high frequency components, followed by zigzag scanning to convert them into one-dimensional data. Finally, DC coefficients are encoded based on Differential Pulse Code Modulation (DPCM) and Huffman encoding, while AC coefficients are encoded using Run-length encoding and Huffman encoding. This process results in the final compressed image.\\
Block quantization processing in DCT transformation can cause blocking to occur when the compression rate is high. In the prior research [1] of this study, an analysis was conducted by converting PNG images into JPEG using different compression rates, which showed that blocking oc-curs at 95\% and 98\% compression rates. The objective of this paper is to focus on image restoration for 98\% compressed JPEG images using multiple loss functions and new network structures. This represents the maximum compression rate, and the resulting compressed images can contain visual content that is unrecognizable. In a preliminary study [2], the image-to-image method was used to determine whether the maximum compressed image could be restored. Figure 1 displays the maximum compressed JPEG image, which exhibits prominent blocking phenomena when observed with the naked eye. Furthermore, significant loss of color information can also occur. Multiple blocking phenomena can lead to degraded image quality and make it impossible to identify faces. If the original image content can be restored from maximum compression, it can be a significant advantage for image security.\\

\begin{figure}[htb!]
    \centering
    \includegraphics[width=15cm]{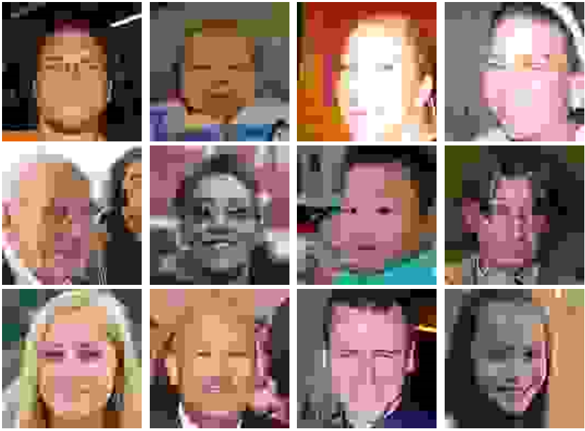}
    \caption{Examples of JPEG lossy maximum-compressed image}
\end{figure}

Our main contribution is as follows.

•	As far as we know, this study is the first attempt to restore images with maximum lossy compression, which is a challenging task in the field of image processing and restoration.\\
•	This method is built upon a generative adversarial network (GAN) and introduces a novel hourglass block at the deep layer through the modification of the U-Net architecture.\\
•	If the image to be restored contains very little information, the performance of the continuous discriminator is found to be poor and is analyzed. To address this issue, this paper proposes using the Discriminator as an Early Stopping method to improve performance.

\section{Related Works}
\label{sec:headings}

\subsection{Lossy Compression}
F. Mentzer et.al.[3] proposed a method of lossless compression using BPG, a lossless image compression algorithm. The original images compressed with BPG and residuals were decomposed into reconstructions to model CNN-based residual distribution and generated results based on resid-ual coders using entropy coding show high compression performance. C. Qin et.al.[4] proposed research on the loss compression method for encrypted images. It showed a method of encrypting using modulo-256, compressing the image and preventing distortion of missing pixels based on im-age inpainting. Z. Yan et.al.[5] proposed a GAN-based framework to produce the result of the low-est MSE distortion at the bit rate. This study shows a theoretical proof that distortion loss does not significantly affect training. N. Gupta et.al.[6] proposed a compression technology using a fractal-based compression algorithm. This shows that the DES encryption system provides efficient recon-struction and security of images based on the minimum bandwidth.

\subsection{Image Generation}
DCGAN[7] can be seen as the beginning of a generation model that improves the vanilla GAN[8]. This model was born after tremendous efforts such as changing the fully-connected struc-ture based on CNN and changing the activation function. This research showed the results of gen-eration of a change in the latent vector called walking in the latent space. Pix2Pix[9] is a GAN-based network created based on cGAN[10]. This is not a random vector, but an image of another style is received as input and output as output. In order to learn this based on supervised learning, data and correct answer images are needed to enter input. PGGAN[11] is a model that increases resolution by starting training from low-resolution images and gradually increasing the size of imag-es by adding layers. Unlike the existing GAN model, it is possible to stably generate high-resolution images as well as shorten learning time by focusing on detailed contents step by step. StyleGAN[12] is a model that changed the structure based on style transfer in the network configuration of PGGAN[11], enabling scale-specific control that PGGAN[11] could not do. This is because the la-tent vector is entangled, which was solved by quantifying it. In addition, AdaIN[13] was used so that one style could affect only each scale. However, defects such as water droplets were found and the higher the resolution, the greater the frequency. The research to solve this problem is StyleGAN2[14]. StyleGAN2[14] solved this problem by referring to normalizing to the mean and variance of each AdaIN[13] as the problem of destroying information between features. StyleGAN-ADA[15], a follow-up study of StyleGAN[12] and StyleGAN2[14], proposed a method to solve the problem of overfitting discriminator during learning through a small amount of datasets. Generator and discriminator were trained using only augmented images, and the augmented range was adjust-ed through probability and solved. These Image Generation studies are actively underway, and there are so many applied studies. It is used in a wide variety of fields such as, human postures based on key points [16], the next frame predicted based on multiple frames [17] and cloud re-moval from solar images [18].

\subsection{Super Resolution}
Although the content pursued in this paper is not technically Super Resolution, it can be said that there is something in common in improving image quality. SRCNN[19] proposed CNN which learns using the end to end mapping method to conduct a study to restore low-resolution images to high-resolution. The structure of SRCNN[19] is characterized by a very simple and fast speed using only convolution layers. ESPCN[20] shows a study that has proposed a sub-pixel layer in order to solve the disadvantage of increasing the computation amount as upscaling is performed at the input of the network. This layer allows the network to calculate as a small image by alternately attaching pixels in the last feature map, and there is an improvement in performance. VDSR[21] proposed a method that can utilize the texture information of the overall image using a deep network using 20 layers. In addition, speed improvement was shown using adjustable gradient clipping. SRGAN[22] has improved image quality by presenting a GAN-based structure to secure the disadvantage that previous studies are difficult to restore information in the high frequency area. Generating a better image for humans using the high-level feature map of the VGG network is similar to the method presented in this paper. ESRGAN[23],  a follow-up study of SRGAN[22] has improved the quality of more natural and realistic images through several methods such as introducing RRDB and using features before activation.

\section{Restoring of Maximum Lossy Compression Image}
\label{sec:headings}
The objective of this paper is to restore maximum lossy compression images, which is a challenging task due to the difficulty of recovering data with limited information as depicted in Fig. 1. However, the proposed GAN-based structure demonstrates high performance in restoring such images to their original state. The overall learning structure of this paper is illustrated in Fig. 2, where the original image is denoted by $x$ and defined as $C(x)$ by generating the maximum lossy compressed image using the JPEG compressor. The supervised learning method is used in the form of a pair between $x$ and $C(x)$, where the training data configuration can be expressed as $S_{data} = {(x_i, C(x_i)) \mid x_i \in X, C(x_i) \in C, i = 1, 2, \ldots, N}$, and the size of the image is $(128, 128, 3)$. The input of the generator is $C(x)$, denoted as $G(C(x))$, which aims to restore the maximum loss compressed image by generating an image that is similar to $x$. The generators aim to create images closer to $x$ to deceive the discriminator, while the discriminator is trained to accurately differentiate between $G(C(x))$ and $x$. The generator and discriminator have a minimax structure that conflicts to generate more natural images, resulting in generated reconstruction images that are closer to the original image.

\begin{figure}[htb!]
    \centering
    \includegraphics[width=13cm]{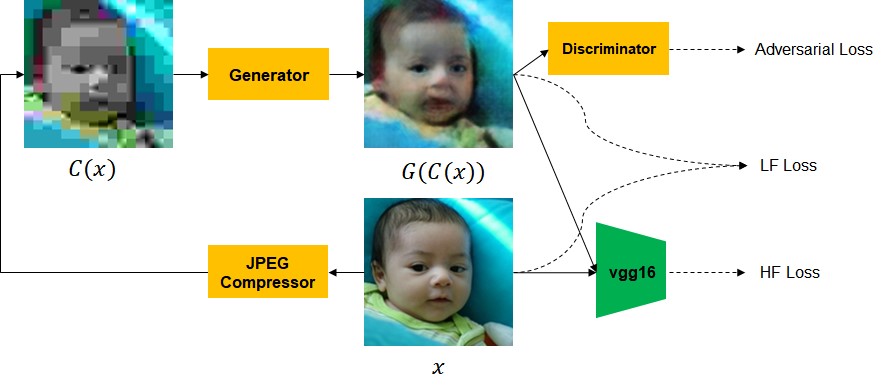}
    \caption{Overall training architecture}
\end{figure}

\subsection{Generator}
The U-Net based network structure of the Generator in this paper, illustrated in Fig. 3, is designed to restore maximum loss compressed images. Unlike the existing U-Net, it has a maximum depth of seven with the input and output designed to be of the same size. The input and output of the network are color images of (512, 512, 3) while the original image size is (128, 128, 3). Encoding extracts features in six layers, and decoding restores them to their original size. Additionally, an hourglass network structure is placed in the middle of the network to extract small features that are lost in the datasets. This structure repeats four times at depths of the 6th and 7th layers to extract useful features, and all layers have a 4x4 filter and stride 2. The depth of the 6th and 7th layers is set to 1024 to extract many features. The activation function for all layers except the last layer of decoding is Leaky-ReLU, and the last layer uses tanh. Finally, the resizing of the decoding result to its original size generates a natural image. However, linear interpolation is used to adjust the image size from (128, 128, 3) to (512, 512, 3), increasing the number of pixels by 16 times, which may result in a longer training time.

\begin{figure}[htb!]
    \centering
    \includegraphics[width=13cm]{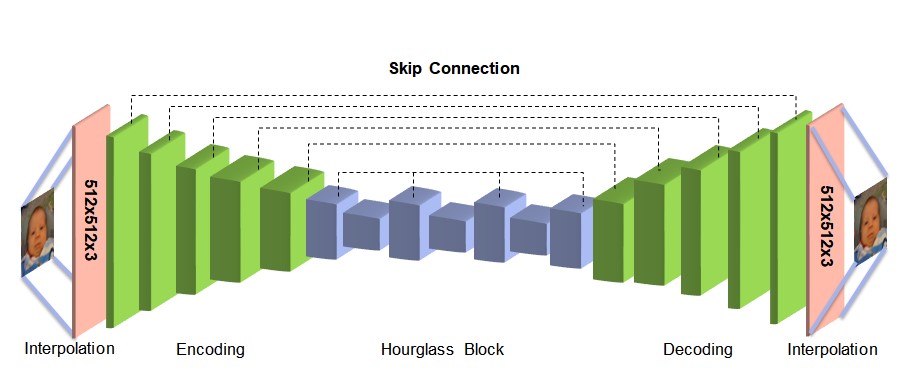}
    \caption{Generator architecture}
\end{figure}

\subsection{Discriminator}

The network structure of the Discriminator in this research is based on the five-layer PatchGAN[23] structure, as shown in Fig. 4. The PatchGAN[23] structure has been demonstrated to improve performance, and in this work, the input of the Discriminator is formed by concatenating $x$ and $G(C(x))$, resulting in a feature map of size (62, 62, 1). This feature map corresponds to the receptive field proposed by PatchGAN[23] and is used to train the Discriminator to classify the generated $G(C(x))$ as Fake. Layers 1 to 3 use Leaky-ReLU as the activation function, and the stride is fixed to 2. In the fourth layer, an activation function using Leaky-ReLU is applied, and the stride is fixed to 1 with padding. In the fifth layer, the activation function uses sigmoid to produce an output between 0 and 1, and the structure is the same as the 4th layer.

\begin{figure}[htb!]
    \centering
    \includegraphics[width=10cm]{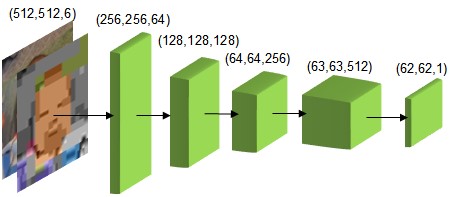}
    \caption{Discriminator architecture}
\end{figure}

\subsection{Loss functions}
The loss function proposed in this paper consists mainly of three types. However, when using the log function commonly used in existing GAN, the Curse of Dimensionality occurs in the proposed method, as explained in Section 4. To achieve stable learning without Curse of Dimensionality, the log function is removed, and the loss is defined solely by the output of the existing Discriminator. The Discriminator is trained to classify $D(G(C(x)))$ as close to 0 and $D(x)$ as close to 1, resulting in an Adversarial loss output value between -1 and 1, which is trained to minimize according to Eq. 1. For the Generator, the aim is to make $G(C(x))$ appear real to the Discriminator. Therefore, the loss is set to minimize Eq. 2. Consequently, Eq. 1 and Eq. 2 have opposite structures and can be used in the GAN training process.

\begin{equation}
\mathcal{L}_{Adv}(D) = \log(D(G(C(x)))) - \log(D(x)) \label{eq1}
\end{equation}

\begin{equation}
\mathcal{L}_{Adv}(G) = \log(1 - D(G(C(x)))) \label{eq2}
\end{equation}

Image are partitioned into two regions: high and low frequency. The high frequency region includes the edges and details of the object, while the low frequency region includes the object's color and texture. In the case of face images, the high frequency region may correspond to the person's identity, while the low frequency region may be referred to as the excluded part. To preserve the low frequency region, the proposed loss function employs the L1 Loss, as shown in Eq. 3, by minimizing the difference between the pixel values of the generated image and the original image. To preserve the high frequency areas, a pretrained VGG-16 model is utilized, and the feature maps that can preserve high frequency regions well are selected using L2 Loss, as defined in Eq. 4, which is referred to as the HF Loss.

\begin{equation}
\mathcal{L}_{LF} = |x - G(C(x))| \label{eq3}
\end{equation}

\begin{equation}
 \mathcal{L}_{HF} = (l_{i}(x)-l_{i}(G(C(x))))^2  \label{eq4}
\end{equation}

The Total loss function, expressed as Eq. 5, is the sum of the three loss functions with the aim of minimizing them. The hyperparameters are set according to the weight of each loss function, as described in Section 4.

\begin{equation}
\mathcal{L}_{total} = \lambda_{Adv} \mathcal{L}_{Adv}(G, D) + \lambda_{LF} \mathcal{L}_{LF} + \lambda_{HF} \mathcal{L}_{HF}  \label{eq5}
\end{equation}

\section{Experiments}
\subsection{Datasets}
The datasets utilized for the training and testing phases were derived from the FFHQ datasets proposed by StyleGAN[11]. FFHQ is a collection of Western facial datasets that contain 4,000 im-ages, which are partitioned into an 8:2 ratio for training and testing purposes. The images are in PNG format, which is lossless and offers superior quality and capacity compared to other formats. To create the maximum lossy compressed images(JPEG compressor), the PNG images were con-verted to JPEG and compressed simultaneously, resulting in a compressed data rate of approximate-ly 96.7\% and an average information retention rate of 3.3\%. The maximum lossy compressed im-ages and the original FFHQ images are color images with a size of (128, 128, 3) and are paired for evaluation.

\subsection{Training Details}
The overall training process utilizes the stochastic gradient descent method and runs for 30 epochs, consisting of 9600 iterations. Both the generator and discriminator employ the Adam optimizer with an initial learning rate of 0.00001. Additionally, dropout with a rate of 70\% is applied to the hourglass block. As the training datasets contain limited information, generating high-quality images is challenging, especially for the generator. As a result, the discriminator cannot effectively distinguish between real and fake images, causing difficulties in producing natural and high-performance videos. To overcome this challenge, the discriminator is not trained after 10 epoch progresses, and a higher quality image is generated. With  $\mathit{\lambda}_{Adv}$ set to 1 and low-frequency images comprising most of the data, $\mathit{\lambda}_{LF}$ is set to a high value of 20. On the other hand, the high-frequency images undergo training with$\mathit{\lambda}_{HF}$ set to 0.1. To detect the high-frequency region, only Conv4\_1 of VGG-16 is used, which preserves the identity of the number of people. Finally, the maximum lossy compressed image is restored, and the generated image is saved in PNG file format, similar to the original image.

\subsection{Results Analysis}
The results for the method proposed in this paper are shown in Fig. 5. The compressed images are located on Fig. 5-(a, d), the original image is located on Fig. 5-(c, f) and the central image(Fig. 5-(b, e)) is the image generated by the generator. Overall, the generator generated an image close to the original image using a compressed image with very little information. In particular, the color of the skin and hair was greatly lost, but it was shown that even this part could be created. However, due to the limitations of the loss-compressed data, the overall shape can be generated with good performance, but it cannot be created for detailed elements. This is because if you look at the com-pressed image of the contents such as wrinkles, whiskers, and hair results of the original image, you cannot find this part even when a person sees it. Because wrinkles cannot be created the generated image represents a young face and it is often found that a closed mouth image shows teeth. This is because there is little information in the compressed data and many of the datasets are smiling, so the generator is created with its mouth open. Therefore, restoring using the maximum lossy com-pressed image has limitations and the method proposed in this paper is located near the limit and can generate a natural and high-performance image.

\subsection{Performance Evaluation}
For image quality evaluation, the performance is assessed based on commonly used metrics such as PSNR and SSIM[24]. Additionally, a further evaluation using VIF[25] is performed.

PSNR is a quality index capable of measuring the loss information of an image due to loss of quality. The better the result, the smaller the difference from a pixel point of view. However, since it only considers the pixel point of view, it can evaluate the low-frequency area but lacks consideration for the human visual aspect. Thus, a high score can be obtained even if a natural image is not generated. The equation for PSNR is presented in Eq. 6, and a higher score corresponds to better performance.

\begin{equation}
{PSNR}(x,G(C(x)) = 10 \cdot \log_{10} \left(\frac{{MAX}^2}{{MSE}}\right)
\label{eq:psnr}
\end{equation}

\begin{figure}[htb!]
    \centering
    \includegraphics[width=14cm]{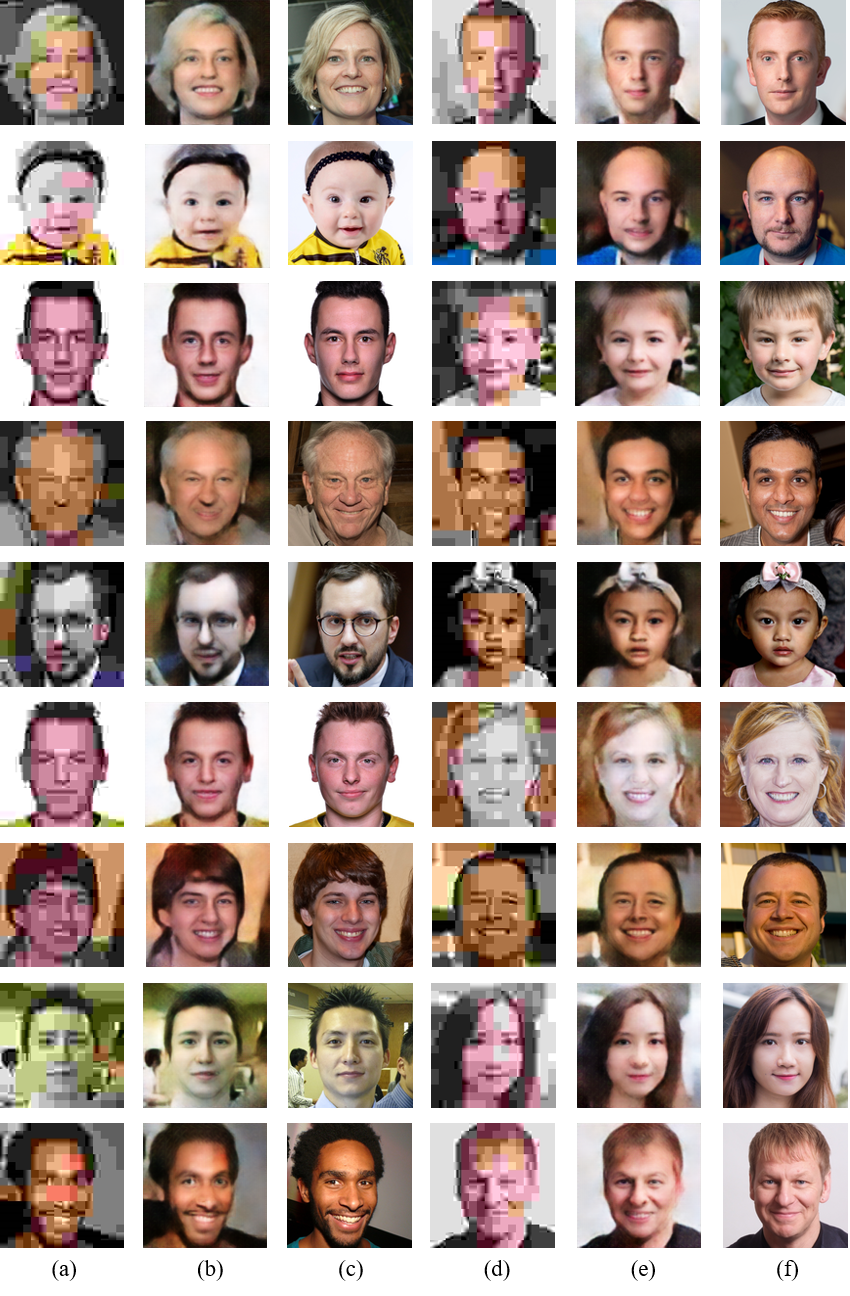}
    \caption{Comparison of Images ((a),(d): Compressed Images, (b),(e): Restored Images, (c),(f): GT)}
\end{figure}

SSIM is a widely used image quality measurement index in engineering that considers human visual content. It evaluates the similarity between an original image and a generated image based on factors such as structure, brightness, and contrast. The closer the SSIM value is to one, the more similar the two images are. Compared to PSNR, which only considers pixel differences, SSIM is a more reliable method as it overcomes the shortcomings of PSNR. SSIM can be calculated using the formula presented in Eq. 7.

\begin{equation}
SSIM(x, G(C(x)) = \frac{({2\mu_x\mu_{G(C(x))} + C_1)} \cdot {(2\sigma_{x,{G(C(x))}} + C_2)}}{({\mu_x^2 + \mu_{G(C(x))}^2 + C_1)} \cdot {(\sigma_x^2 + \sigma_{G(C(x))}^2 + C_2)}}
\label{eq:ssim}
\end{equation}

VIF[25] is a metric that measures the fidelity of an entire reference image to evaluate its quality. It does so by comparing the information content between the original and generated images. This indicator uses the results of the human visual system and Gaussian distribution to assess the quality of the image. A high information content in the generated image, even after being input as a lossy compressed image, is indicative of good performance. The evaluation of performance is expressed, where a VIF value closer to one indicates higher performance.

The study compares the generation results of previous researches with the proposed method in Fig. 6, and presents a variety of performance comparisons in Table 1. The proposed method shows high performance compared to the previous researches. The study additionally evaluates the performance of waifu2x[26] and Pix2Pix [8], which improve the resolution of CNN-based 2D images. $C(x)$ refers to the lossy compressed image that exhibits blocking artifacts in Fig. 6-(a), but after applying waifu2x[26] (Fig. 6-(b)), there is no significant difference from the original image. Although super-resolution methods are generally necessary for evaluating the performance of the study, the purpose of the research is to restore the lossy compressed image rather than to improve its quality. Therefore, the existing image quality improvement methods do not show a significant effect. Pix2Pix[8] (Fig. 6-(c)) shows better performance than waifu2x[26], but details such as eyes, nose, and mouth do not appear natural. The proposed method in Fig. 6-(d) and 6-(e) produces more natural and clearer results than the previous researches.

\begin{table}[htbp]
 \caption{Performance comparison including maximum lossy compressed images}
  \centering
  \begin{tabular}{lrrrr}
    \toprule
     Metrics     & Lossy Compress.     & Pix2Pix[8] &waifu2x[26] &Ours \\
     \midrule
    PSNR &20.381 &20.537 &21.181 & \textbf{21.680} \\
     \midrule
    SSIM & 0.582 & 0.586 &0.561 & \textbf{0.675} \\
     \midrule
    VIF & 0.201 & 0.211 &0.181 & \textbf{0.261} \\

    \bottomrule
  \end{tabular}
  \label{tab:table}
\end{table}

\subsection{Ablation Study}
The concept of GAN is a critical aspect of this research since it enhances the performance of the generator and discriminator modules compared to using only Convolutional Neural Network. The ultimate goal of GAN is for the generator to produce highly realistic data and for the discriminator to distinguish between real and generated images. However, the generator operates based on the maximum lossy compressed image, which makes it impossible to generate a realistic image that can fool the discriminator. This is due to the minimal amount of information in the maximum lossy compressed image, which makes it easy for the discriminator to distinguish between real and generated images. Although the GAN structure may work effectively in the early stages before the discriminator is trained, the generator's focus on beating the discriminator, rather than training for good performance, prevents it from generating high-quality images. This research proposes a new training method in which the discriminator is trained for only 10 epochs, after which the generator is trained to optimize a loss function other than beating the discriminator. Table 2 presents the results of comparative analysis to determine whether to change $\mathit{\lambda}_{LF}$, which is significantly involved in generating low frequencies and attracting the discriminator to 10 epochs. The findings show that $\mathit{\lambda}_{LF}$ achieves the highest performance at 20, and stopping the discriminator's training in all cases leads to improved results.\\
The proposed contribution in this paper is assessed by evaluating the performance of the model with and without the hourglass block structure and VGG-16. The results indicate that without the hourglass block structure and VGG-16, the performance is low in all aspects compared to the high-est performance. The method proposed in this paper is a slightly improved version of Pix2Pix[8], with the network and loss function being fixed and only linear interpolation being added in the in-put. The addition of the hourglass block structure resulted in an improvement in performance, demonstrating that the block can aid in maximizing the representation of the conserved portion of the high-dimensional feature map. Subsequently, the addition of high-frequency region preservation using VGG-16 resulted in achieving the highest performance mentioned in this paper. The pro-posed method not only preserves the natural images but also maintains the identity of the person, resulting in a relatively high-quality image.

\begin{figure}[htb!]
    \centering
    \includegraphics[width=15cm]{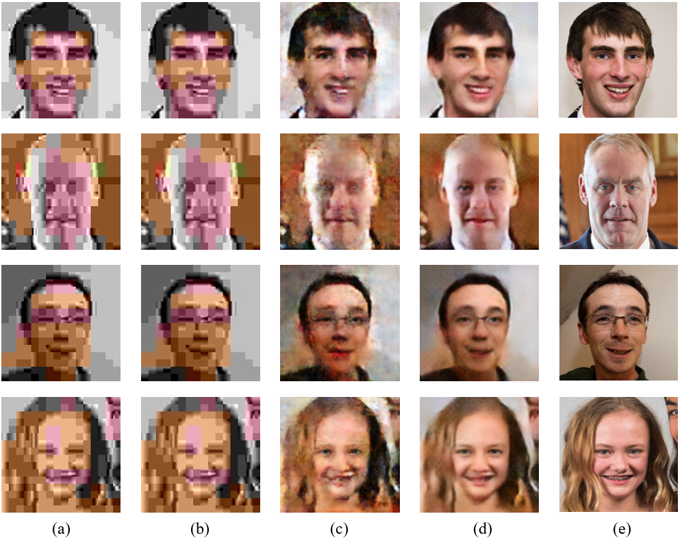}
    \caption{Comparison of previous researches}
\end{figure}

\begin{table}[htb!]
 \caption{The results of the experiment according to not stop or stop of discriminator training and hyperparameter}
  \centering
  \begin{tabular}{|c|c|c|c|c|c|c|}
    \hline
    {Metrics} & \multicolumn{2}{c|}{$\mathit{\lambda}_{LF}$ =5} & \multicolumn{2}{c|}{$\mathit{\lambda}_{LF}$ =10} & \multicolumn{2}{c|}{$\mathit{\lambda}_{LF}$ =20}\\
    \cline{2-7}
     & nonstop & stop  & nonstop & stop  & nonstop & stop\\
    \hline
     PSNR & 21.558 & 21.617 & 21.573 & 21.675 & 21.596 & \textbf{21.680} \\
    \hline
    SSIM & 0.666 & 0.670 & 0.670 & 0.673 & 0.671 & \textbf{0.675} \\
    \hline
    VIF & 0.255 & 0.257 & 0.257 & 0.259 & 0.257 & \textbf{0.261} \\
    \hline
  \end{tabular}
  \label{tab:table}
\end{table}

\begin{table}[htb!]
 \caption{Comparison of performance according to the use of hourglass block and vgg-16}
  \centering
  \begin{tabular}{|c|c|c|c|}
    \hline
     Hourglass Block & X & O & O \\
     VGG-16 & X & X & O \\
    \hline
    PSNR  & 20.623 & 21.445 & \textbf{21.680} \\
    \hline
    SSIM  & 0.604 & 0.665 & \textbf{0.675} \\
    \hline
    VIF  & 0.222 & 0.255 & \textbf{0.261} \\
    \hline
  \end{tabular}
  \label{tab:table}
\end{table}

\subsection{Limitations}
Due to the focus of this paper on restoration based on the maximum compression loss image, the restoration process is limited in its ability to produce exact replicas. However, the objective is to create datasets that closely resemble the original through deep learning network-based learning. While some restoration is feasible at the general shape level, the details cannot be restored. Although similar colors can be generated, wrinkles and hair features are not reproduced. Moreover, the restoration performance is inferior for the datasets on the side of the face, as demonstrated in Fig. 7. This is due to the fact that the majority of the datasets used for restoration have a forward-facing orientation, and the restoration technique restores the missing content based on its characteristics. As shown in the restored image in Fig. 7, the opposite eye is generated despite not having any relation to the frontal data. Additionally, the face shape is blurred. In order to improve the restoration performance of the side images, it is necessary to include side-facing images in the training data.

\begin{figure}[htb!]
    \centering
    \includegraphics[width=10cm]{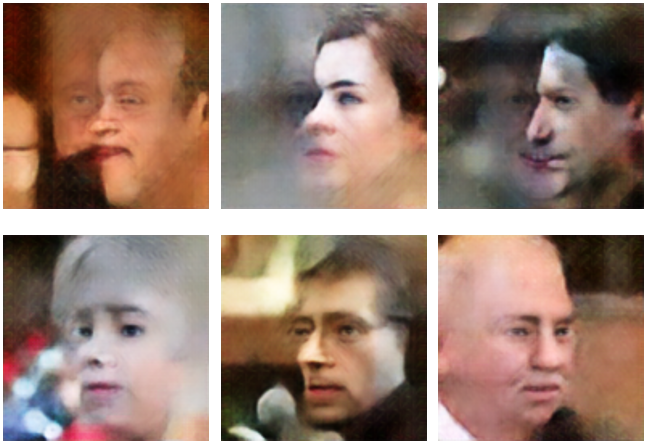}
    \caption{The results of the face side image}
\end{figure}

\section{Conclusions}
This paper presents a GAN-based network approach for restoring JPEG images using data with maximum lossy compression. Despite the difficulty of restoring images with minimal information, even to the human eye, the proposed method demonstrates satisfactory performance. Although there are some limitations to achieving precise restoration, the method has shown that it can restore images to some degree. Future work is expected to improve restoration performance further, poten-tially expanding the application of lost compressed image restoration to the field of image security.

\bibliographystyle{unsrt}

\end{document}